\documentclass[
twocolumn,
aps,showpacs,superscriptaddress,superscriptaddress,
amsthm,amsmath,amssymb
]{revtex4-1}   
\usepackage{graphicx}
\usepackage{dcolumn}
\usepackage{bm}
\usepackage{hyperref}
\usepackage{mathtools}
\usepackage{color}
\usepackage{enumitem}
\usepackage{chemformula}
\usepackage{amssymb,amsmath,amsthm,amsfonts}
\usepackage[caption=false,justification=justified]{subfig}  
\usepackage[normalem]{ulem}
\usepackage{chemfig}
\usepackage[braket]{qcircuit}

\DeclareMathOperator\Tr{Tr}

\begin{document}

\global\long\def\U{\mathbf{U}}
\global\long\def\l({\left(}
\global\long\def\r){\right)}


\global\long\def\lc{\left\{}
\global\long\def\rc{\right\}}
\global\long\def\R#1{R_{\left|#1\right\rangle }}
\global\long\def\op#1#2{\left|#1\right\rangle \left\langle #2\right|}
\global\long\def\ip#1#2{\left\langle #1,#2\right\rangle }
\global\long\def\s#1{\sum_{#1\in\left\{  0,1\right\}  ^{k}}}
\global\long\def\I{\mathbf{I}}
\global\long\def\ve{\varepsilon}

\newcommand{\eq}[1]{(\ref{eq:#1})}
\renewcommand{\sec}[1]{\hyperref[sec:#1]{Section~\ref{sec:#1}}}
\newcommand{\fig}[1]{\hyperref[fig:#1]{Fig.~\ref{fig:#1}}}
\newcommand{\tab}[1]{\hyperref[tab:#1]{Table~\ref{tab:#1}}}
\newcommand{\routine}[1]{\hyperref[#1]{Routine~\ref{#1}}}

\newcommand{\Gate}[1]{\textsc{#1}}
\newcommand{\hgate}{\Gate{h} }
\newcommand{\zgate}{\Gate{z}}
\newcommand{\tgate}{\Gate{t} }
\newcommand{\pgate}{\Gate{p}}
\newcommand{\notgate}{\Gate{not} }
\newcommand{\cnotgate}{\Gate{cnot} }
\newcommand{\rzgate}{{\Gate{r}}_{z}}
\newcommand{\rygate}{{\Gate{r}}_{y} }
\newcommand{\rxgate}{{\Gate{r}}_{x}}
\newcommand{\gmsgate}{{\Gate{gms}}}
\newcommand{\czgate}{{\Gate{c-z}}}
\newcommand{\braket}[2]{{\left< #1  \left.\right| #2\right>}}

\newcommand{\GF}{\operatorname{GF}}
\DeclarePairedDelimiter\norm{\|}{\|}

\newcommand{\nam}[1]{\textbf{\color{red}[Nam: #1]}}
\newcommand{\stefan}[1]{\textbf{\color{blue}[Stefan: #1]}}


\newcommand{\comment}[1]{}

\title{A Quantum Algorithm for Network Reliability}

\author{Stefan Pabst}
\affiliation{RelationalAI, 2120 University Avenue, Berkeley, CA 94704, USA}
\author{Yunseong Nam}
\affiliation{Department of Physics, University of Maryland, College Park, MD 20742, USA}

\date{\today}

\begin{abstract}
Building a network that is resilient to a component failure is vital. Our access to electricity and telecommunications or the internet of things all hinge on an uninterrupted service provided by a robust network. Calculating the network reliability $R$
is $\sharp$P-complete and intractable to calculate exactly for medium and large networks.
Here, we present an explicit, circuit-level implementation of a quantum algorithm that computes $R$.
Our algorithm requires $O(EV/\epsilon)$ gate operations and $O(E)$ qubits, where $V$ and $E$ are the number of nodes and edges in the graph and $\epsilon$ is the uncertainty in the reliability estimation.
This constitutes a significant polynomial speedup over the best classical approaches currently known.
We further provide quantum gate counts, relevant for both pre-fault-tolerant and fault-tolerant regimes, sufficient to compute $R$. 
\end{abstract}
 
\maketitle

\section{Introduction}
\label{sec:introduction}

Moore and Shannon~\cite{MooreShannon1956} showed in 1956 that network reliability can be stated as a probabilistic model and combining networks into a larger overall network
can improve the overall reliability of the network~\cite{Rose2018Overview}.
Ever since, studying the computation of various aspects of network reliability became an active research topic, and has only increased in its importance;
From electric grids and internet to any kind of transportation system, networks are ubiquitous in our daily lives. 
We depend on their reliability, to prevent electric outages or to ensure stable internet connections.  

Computing the reliability, though, is known to be a computationally hard problem that is $\sharp$P-complete~\cite{Ball86-Overview, Valiant1979}, making its incorporation into a design criterion difficult despite its importance~\cite{Handbook, OsorioVardi2017}.
More specifically, network reliability $R$ is defined as the probability that, if each edge $\varepsilon$ fails with a probability $p_{\varepsilon}$, the remaining graph is connected.
For a network $G({\mathcal V},{\mathcal E})$ with a set of nodes ${\mathcal V}$ and a set of edges ${\mathcal E}$, the network reliability can be calculated as~\cite{Rose2018Overview, HuiPhD2005}
\begin{align}
\label{eq:network-reliability}
R = \sum_{{\mathcal E}' \subseteq {\mathcal E}} \left(\prod_{\varepsilon \in {\mathcal E}'} p_{\varepsilon}  \prod_{\varepsilon \notin {\mathcal E}'} (1-p_\varepsilon)\right),
\end{align}
where ${\mathcal E}'$ is a subset of ${\mathcal E}$ such that the subgraph $G'=({\mathcal V}, {\mathcal E}')$ is connected.
Naively enumerating all $2^{E}$ unique subsets ${\mathcal E}'$ and checking whether the corresponding subgraph $G'$ is connected requires an exponentially large number of computational steps in the network size, rendering the exact computation impractical~\cite{BALL1995673,4298237,52639,210267}.
Here, $E=|{\cal E}|$ is the cardinality of the set ${\cal E}$ and corresponds to the number of edges in the graph.
Similarly, the number of nodes is then given by $V=|{\cal V}|$.

There exist multiple variants of network reliability problems depending on whether or not the edges are directed or undirected and the number of terminal nodes $T$ ($T$-terminal) that need to be connected~\cite{GertsbakhBook2011, Rose2018Overview, HuiPhD2005}.
We refer the readers to Refs.~\cite{GertsbakhBook2011, Rose2018Overview, Handbook,6903608, PAREDES2019106472} for an overview of the network reliability activity.

Briefly, computational methods calculating the network reliability can be categorized into three groups: exact or bounds, guarantee-less simulation, and probably approximately correct (PAC)~\cite{PAREDES2019106472}.
Exact methods, such as the naive enumeration, are feasible for small graphs.
Common approximate methods are Monte-Carlo based~\cite{PAREDES2019106472, VAISMAN20161,10.5555/1823390} or use binary decision diagrams~\cite{6903608,4220790}.
These methods often lack a rigorous error analysis
rendering these methods \textit{guarantee-less} in terms of correctness~\cite{PAREDES2019106472}.
Complexity-wise, the best method we are aware of for an all-terminal problem runs in time $O(E^2V^3/\epsilon^2)$~\cite{FPRAS}.

In practice, a worse-scaling, yet more utilitarian approach can be employed, including a method, extensible to $T$-terminal problems, that makes $O(\log(v)\log(1/\delta)/\epsilon^2)$ calls to an \textit{oracle} (SAT solver).
The oracle solves a non-deterministic polynomial time decision problem with a number of variables $v$ that scales as $O(V+E\log(E/\epsilon))$~\cite{PAREDES2019106472}.
Note that each oracle call has its own inherent complexity.
Here, $\epsilon$ is the uncertainty in the reliability estimation and $\delta$ is the confidence in the correctness of the reliability. Note the $\delta$-dependence of the latter method makes the approach probably approximately correct.

In this paper, we show a quantum algorithm for computing the network reliability $R$ for an undirected graph.
Thanks to the quantum superposition, we can realize all subgraphs $G'$ simultaneously, check they are connected, and count the outcomes.
Our proposed algorithm is deterministically approximately correct and independent of $\delta$.
It is fully compatible with an arbitrary $T$-terminal problem, admits a gate complexity of $O(EV/\epsilon)$, and the number of qubits used is $O(E)$. 

Our paper is organized as follows. In \sec{outline}, we provide an outline of our algorithm.
In \sec{algorithm}, we provide the details of the algorithm, including the explicit circuits written in the standard, elementary gate set of $\cnotgate$ and single qubit gates for all necessary oracles. We leverage in-circuit measurements, used to effectively act an OR operation on a quantum computer, to keep the number of ancilla qubits minimal.
In \sec{complexity}, we calculate the overall complexity of our algorithm.
In \sec{gate-count}, we provide an estimate for gate counts in terms of \cnotgate and \tgate gates, useful for pre-fault tolerant and fault tolerant regimes, respectively.
We conclude our paper in \sec{conclusion} and discuss the implications of our results.

\section{Outline of the Algorithm}
\label{sec:outline}

In this section, we outline the steps of our algorithm. 
The exact implementation details, including the circuit construction methodologies and circuit element labels, are presented in \sec{algorithm}.

The key idea of our algorithm is to exploit quantum superposition to create all possible edge-failure configurations ${\cal E}'$, apply a reachability computation, and then measure its outcome for all the edge-failure configurations at once.
The reachability computation follows a simple idea where we start at an arbitrarily chosen \textit{root} node and we check that we can reach all other nodes starting from this node.
Because we are dealing with an undirected graph, calculating the reachability from one arbitrarily chosen node to all other nodes is equivalent to calculating the all-terminal network reliability~\cite{Ball1980ComplexityON}.
As we will see, our approach can be straightforwardly extended to a $T$-terminal reliability problem for undirected graphs.

To realize this reachability computation on a quantum computer, we mainly use two quantum registers, namely, edge qubits $e_\varepsilon$ with $\varepsilon\in\{0,1,..,E-1\}$ and node qubits $v_i$ with $i\in\{0,1,..,V-1\}$.
We abuse the notations throughout the paper, where $\varepsilon$ refers to both the edges $(i, j)$, connecting the nodes $v_i$ and $v_j$, and their enumerations. 
Similarly, we use a double index for the edge qubits $e_{i,j}$, when we want to stress that the edge is connecting the nodes $v_i$ and $v_j$.
As will be shown later in our detailed implementation, we additionally use a single ancilla qubit and a label qubit.
Each qubit in the edge register encodes the failure probability of each edge $\varepsilon \in {\cal E}$.
Each qubit in the node register encodes the reachability of the node from the \textbf{root} node.
Specifically, our algorithm performs the following steps:

\subsection{Step 1}
We initialize the edge register to
\begin{align}
    \label{eq:edge-preperation}
    \ket{\psi_\text{edge}} 
    &= \bigotimes_{\varepsilon=0}^{E-1} \left(\sqrt{p_{\varepsilon}} \ket{e_\varepsilon=0} + \sqrt{1-p_{\varepsilon}} \ket{e_\varepsilon=1} \right)
    \\
    &= \sum_{e = 0}^{2^E-1} f_e \ket{e}, \nonumber
\end{align}
where $p_\varepsilon$ is the probability of failure of the $\varepsilon$-th edge, $e$ is a number in the binary basis that encodes the bitstring $e_0, \ldots, e_{E-1}$ and corresponds to a specific computational basis state of the edge register.
The bit $e_\varepsilon$ corresponds to the computational basis state of the $\varepsilon$-th qubit in the edge register, and specifies whether or not the edge $\varepsilon$ has failed.
The number $e$, therefore, refers to a specific edge failure configurations---for which we have $2^E$ possibilities.

Each edge failure configuration $e$ corresponds to one specific subset $\cal E'$ in (\ref{eq:network-reliability}), where $e_\varepsilon=1$ if $\varepsilon \in \cal E'$ and $e_\varepsilon=0$ otherwise.
The coefficient $f_e$ is the amplitude of the edge-failure state $\ket{e}$ and its absolute square corresponds to the probability that this edge failure configuration $e$ occurred, i.e.,
\begin{align}
   \label{eq:edge-failure-probability}
   |f_e|^2 &= \prod_{\varepsilon \in \{k | e_k=0\}} p_{\varepsilon}  \prod_{\varepsilon \in \{k | e_k=1\}} (1-p_\varepsilon).
\end{align}

\subsection{Step 2}
We initialize the node register to
\begin{align}
    \ket{\psi_\text{node}} 
    &=
    \ket{v_0=1}
    \otimes \left(
    \bigotimes_{i=1}^{V-1} \ket{v_i=0}
    \right)
    , \nonumber 
\end{align}
where, by convenience, $v_0$ is the arbitrarily chosen \textit{root} node which is initialized in state $\ket{v_0 = 1}$ and all other node qubits $v_i$ are initialized in state $\ket{v_i=0}$ indicating that they have not been reached yet.

\subsection{Step 3}

We calculate simultaneously for all edge-failure configurations $\ket{e}$ whether or not the graph is connected.
As stated, this in our case is equivalent to a reachability problem starting from a \textit{root} node.
The main strategy is to check for each node whether it is root-connected, determined by checking if its neighbors are root-connected and the edges to the neighbors have not failed.
This iteration we call the \textit{inner} loop and denote as ${\mathcal C}_\text{inner}$.
The \textit{outer} loop is denoted as ${\mathcal C}$ and is the repetition of the inner loop $V-1$ times, 
$${\mathcal C} = ({\mathcal C}_\text{inner})^{V-1},$$  because any shortest path between two nodes in an undirected graph involves no more than $V-1$ edges.

\subsubsection{Step 3a}
The outer loop induces
\begin{align}
   \label{eq:outer-loop}
   {\mathcal C}\Big(\ket{\psi_{\rm edge}} \ket{\psi_{\rm node}} \Big) 
   & =
   \sum_{e=0}^{2^E-1} (-1)^{\alpha_e} f_e \ket{e}\ket{n_e},
\end{align}
where initially the edge and node registers are independent but get entangled with each other as we apply quantum operations, and $(-1)^{\alpha_e}$ is the sign of the coefficient and depends on the edge-failure configuration $e$ and the specific iteration order (see Sec.~\ref{sec:reachability} for details). 
The final node register state $\ket{n_e}$, to this end, contains the reachability between the root node and every other node and depends on $e$.
The edge-failure configuration $e$ corresponds to a connected graph, if we find the node register in the state 
$$\ket{n_e} = \bigotimes_{i=0}^{V-1} \ket{v_i=1}.$$

\subsubsection{Step 3b}
In the inner loop, we go over all edges $\varepsilon = (i,j) \in \cal E$ and determine that node $j$ is reachable from the root. The node is reachable if it is already determined to be reachable or one of the neighboring nodes is reachable and the edge between them has not failed.
This can be expressed with 
\begin{subequations}
\begin{align}
  {\cal C}_\text{inner} 
  &= \prod_{(i,j)\in \cal E} {\cal C}^{(i,j)}_\text{inner}, 
  \\
  {\cal C}^{(i, j)}_\text{inner} 
  &=
  {\cal C}^{(i, j)}_\text{qc-OR}\, {\cal C}^{(j, i)}_\text{qc-OR},  
  \\
  {\cal C}^{(i,j)}_\text{qc-OR} \ket{e_{i,j}} \ket{v_{i}, v_{j}} 
  &\rightarrow \ket{e_{i,j}}\,\ket{ v_{i}, v_{j} {\oplus} (v_{i} {e}_{i,j})\bar{v}_{j}},
\end{align}
\end{subequations}
and $i \in {\cal V}_j$ are the neighboring nodes connected to node $j$ via the edge $(i, j)$, and ${\cal V}_j = \{i \, | \, (i, j) \in {\cal E}\}$ is the set of all nodes that are directly connected to node $j$.

The expression $v_{j} \oplus (v_{i} {e}_{i,j})\bar{v}_{j}$ is equivalent to the logical expression $v_j \vee (v_i \wedge e_{i, j})$, where $\bar{v}_{j}$ is the negation of the bit $v_j$.
It is the OR of $v_j$ and $v_i$, computed conditionally on $e_{i,j}$ (hence the name quantum-controlled OR or qc-OR). 
It evaluates to 1 if node $j$ is already reachable ($v_j=1$) or node $j$ is not reachable yet ($v_j=0$) but the neighboring node $i$ is already reachable ($v_i=1$) and the edge $(i,j)$ did not fail ($e_{i,j}=1$).
It computes to 0 otherwise.
In Sec.~\ref{sec:reachability}, we discuss in detail how this logic is implemented on a quantum computer.

\subsection{Step 4}
We now determine the reliability. 
Once the quantum state is prepared from the previous step, we can estimate the probability of observing a particular computational outcome, i.e., network reliability.
This is conceptually best done by creating a label qubit $\ket{\lambda}$, which is initialized to $\ket{0}$ and is then set to $\ket{1}$ if all nodes are reachable; The node register in this case is in state $\ket{v_0=1,v_1=1, .., v_{V-1}=1}$. Thus, we induce
\begin{align}
   \ket{n_e} \ket{0}
   &\rightarrow \ket{n_e} \ket{v_0 v_1\ldots v_{V-1}},
\end{align}
which, applied to the entire quantum state, results in
\begin{align}
   \label{eq:label-qubit}
   \sum_{e=0}^{2^E-1} f_e \ket{e} \ket{n_e} \ket{\lambda=0}
   & \rightarrow
   \sum_{e=0}^{2^E-1} f_e \ket{e} \ket{n_e} \ket{\lambda_e},
\end{align}
where $\lambda_e$ indicates that the state of the label qubit is solely dependent on the edge-failure configuration $e$ and it encodes the information on whether or not $e$ represents a connected graph.

By repeatedly performing direct measurements to collect statistics on the label qubit's probability to be measured in $\ket{1}$, we can estimate network reliability $R$.
Specifically, if we label the final state of the quantum registers as $\ket{\psi_\text{final}}$, we can write the reliability as
\begin{align}
    \label{eq:measurement}
    R &= \Tr\left[\mathbb{P}_{\lambda=1} \, \rho_\text{final} \right] 
    = \sum_{e = 0}^{2^{E}-1} |f_e|^2\, \delta_{\lambda_e, 1}
\end{align}
where $\ket{\psi_\text{final}} = \sum_{e=0}^{2^E-1} (-1)^{\alpha_e} f_e \ket{e} \ket{n_e} \ket{\lambda_e}$, the density matrix of the final state is $\rho_\text{final} = \ket{\psi_\text{final}}\bra{\psi_\text{final}}$, and $\mathbb{P}_{\lambda=1} = \ket{\lambda=1}\bra{\lambda=1}$ is the projection operator on the subspace where the label qubit $\lambda$ is in state $\ket{1}$.
An interesting fact is that $R$ does not depend on the phase factor $(-1)^{\alpha_e}$ because the reliability is an incoherent sum of probabilities over all edge configurations.
By inserting (\ref{eq:edge-failure-probability}) in (\ref{eq:measurement}), we arrive at the original definition of network reliability stated in (\ref{eq:network-reliability}).

Note, this can be done more efficiently than repeated measurements, by using the generalized Grover search or amplitude amplification~\cite{brassard2002quantum}, isolating the state with $v_l=1$ for all $l$.
In Sec.~\ref{sec:oracle}, we describe this step in more detail.

\section{Circuit Construction}
\label{sec:algorithm}

\begin{figure}
\[
\Qcircuit @C=0.6em @R=1.0em {
&&&& \lstick{v_0}  & \qw & \multigate{3}{\text{Init}_\text{node}}                  & \qw & \multigate{10}{\mathcal C} & \qw\\
&&&& \lstick{v_1}  & \qw & \ghost{\text{Init}_\text{node}} & \qw & \ghost{\mathcal C} & \qw \\
&&&& \vdots        &     &                        & \vdots & &  \\
&&&& \lstick{v_{V-1}} & \qw & \ghost{\text{Init}_\text{node}}& \qw & \ghost{\mathcal C} & \qw
 \inputgroupv{1}{4}{0.5em}{2.4em}{\ket{\psi_\text{node}}\:} \\
\\
&&&& \lstick{e_0}  & \qw & \multigate{3}{\text{Init}_\text{edge}} & \qw & \ghost{\mathcal C} & \qw\\
&&&& \lstick{e_1}  & \qw & \ghost{\text{Init}_\text{edge}}        & \qw & \ghost{\mathcal C} & \qw\\
&&&& \vdots        &     &     & \vdots & & \\
&&&& \lstick{e_{E-1}} & \qw & \ghost{\text{Init}_\text{edge}}     & \qw & \ghost{\mathcal C} & \qw
\inputgroupv{6}{9}{0.5em}{2.4em}{\ket{\psi_\text{edge}}\:\:\:\:} \\
\\
&&&& \lstick{\ket{\psi_\text{aux}=0}}  & \qw &  \qw & \qw & \ghost{\mathcal C} & \qw \\
}
\]
    \caption{\label{fig:StatePrep} Circuit diagram for the state preparation. Both edge and node register qubits are assumed to be in an all-zero state at the beginning.}
\end{figure}
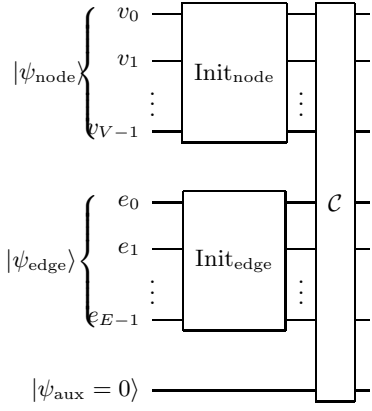

In this section, we lay out the detailed implementation of the algorithm we outlined above.
Specifically, we detail the transformations and registers used to implement the algorithm.
A high-level circuit that implements our algorithm is presented in Fig.~\ref{fig:StatePrep}.

We start with the discussion of the initial preparation of node and edge registers in \sec{register-preparation}.
One of the central transformations to be used in our algorithm thereafter is the reachability operator $\mathcal{C}$.
We present the details of its construction in \sec{reachability}.
We also present the construction of the necessary Grover oracle in \sec{oracle} for completeness.

\subsection{Node and Edge Register Preparations}
\label{sec:register-preparation}

\begin{figure}
\[
\Qcircuit @C=0.6em @R=1.45em {
&&&& \lstick{e_0}  & \qw & \multigate{3}{\text{Init}_\text{edge}} & \qw \\
&&&& \lstick{e_1}  & \qw & \ghost{\text{Init}_\text{edge}}        & \qw \\
&&&& \vdots        &     &     & \vdots & & \\
&&&& \lstick{e_{E-1}} & \qw & \ghost{\text{Init}_\text{edge}}     & \qw \\
}
 \raisebox{-2.5em}{\hspace{4mm}=\hspace{4mm}}  
\Qcircuit @C=0.6em @R=1.0em {
&&&& \lstick{e_{0}}   & \qw & \qw & \gate{\rygate(\theta_0)} & \qw & \qw & \qw \\
&&&& \lstick{e_{1}}   & \qw & \qw & \gate{\rygate(\theta_1)} & \qw & \qw & \qw \\
&&&& \vdots           & & & & \vdots & & \\
&&&& \lstick{e_{E-1}} & \qw & \qw & \gate{\rygate(\theta_{E-1})} & \qw & \qw & \qw  \\
}
\]
    \caption{\label{fig:EdgeRegister} Circuit diagram for the edge register initialization. The circuit implements the state initialization step detailed in (\ref{eq:initialization}) for the edge register. We assume the quantum computer is initialized to an all-zero state prior to quantum gate applications.}
\end{figure}
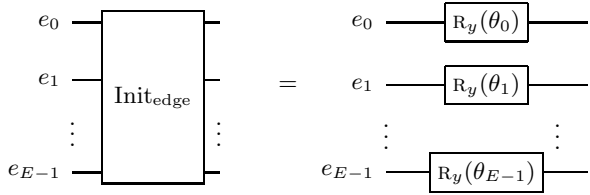

To encode the probability of edge failure, we use a quantum register of size $E$ qubits, where $E$ is the number of edges of the input graph $G$. This can be done by applying $\rygate(\theta):=\exp(-i\theta \sigma_y /2)$ gates ($\sigma_y$ is the Pauli-$y$ operator) on a quantum register, initialized as $\ket{0}^{\otimes E}$ (See Fig.~\ref{fig:EdgeRegister}).
The edge register state is then 
\begin{align}
    \ket{\psi_{\rm edge}}
    &=
    \bigotimes_{\varepsilon=0}^{E-1}
    \rygate(\theta_\varepsilon) \ket{e_\varepsilon=0} \nonumber \\
    &=
    \bigotimes_{\varepsilon=0}^{E-1} \left(\sqrt{p_{\varepsilon}} \ket{0} + \sqrt{1-p_{\varepsilon}} \ket{1} \right),
    \label{eq:initialization}
\end{align}
where $\theta_\varepsilon = \cos^{-1}(\sqrt{p_\varepsilon})$ so that $\cos^2(\theta_\varepsilon) = p_\varepsilon$, the probability of edge failure.
This way, the probabilities of the edge register being in a particular state $\ket{e_0, \ldots, e_{E-1}}$ is given by (\ref{eq:edge-failure-probability}).

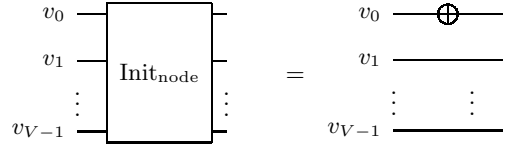
\begin{figure}
\[
\Qcircuit @C=0.6em @R=1.0em {
&&&& \lstick{v_0}  & \qw & \multigate{3}{\text{Init}_\text{node}} & \qw \\
&&&& \lstick{v_1}  & \qw & \ghost{\text{Init}_\text{node}}        & \qw \\
&&&& \vdots        &     &     & \vdots & & \\
&&&& \lstick{v_{V-1}} & \qw & \ghost{\text{Init}_\text{node}}     & \qw \\
}
 \raisebox{-2.5em}{\hspace{4mm}=\hspace{4mm}}  
\Qcircuit @C=0.6em @R=1.45em {
&&&& \lstick{v_{0}}   & \qw & \qw & \targ & \qw & \qw & \qw \\
&&&& \lstick{v_{1}}   & \qw & \qw & \qw & \qw & \qw & \qw \\
&&&& \vdots           & & & & \vdots & & \\
&&&& \lstick{v_{V-1}} & \qw & \qw & \qw & \qw & \qw & \qw  \\
}
\]
    \caption{\label{fig:NodeRegister} Circuit diagram for the node register initialization. Again, as in the edge register, we assume the quantum computer is initialized to an all-zero state prior to quantum gate applications.}
\end{figure}

Encoding the initial node reachability is straightforward. Once we prepare an all-zero classical state, apply a \notgate gate to an arbitrarily chosen qubit, e.g., $l=0$, so that
\begin{align}
  \ket{\psi_{\rm node}} 
  &=
  \notgate \ket{v_0 = 0} \otimes
  \left(\bigotimes_{l=0}^{V-1} \ket{v_l=0}\right)
  \nonumber \\
  & =
  \ket{1} \otimes \left(\bigotimes_{l=1}^{V-1} \ket{v_l=0}\right).
\end{align}
Figure~\ref{fig:NodeRegister} shows the explicit circuit diagram for this node-register initialization step.

\subsection{Reachability Operator}
\label{sec:reachability}

In this subsection, we explicitly construct a circuit that implements the reachability operator ${\mathcal C}$. We describe the operator in two steps: An inner loop, ${\mathcal C}_\text{inner} = \prod_{\varepsilon}{\mathcal C}^{(\varepsilon)}_\text{inner}$, traverses over every edge $\varepsilon=(i,j) \in \cal E$ and applies the operation ${\mathcal C}^{\varepsilon = (i,j)}_\text{inner}$
and an outer loop that repeats the inner loop $V-1$ times, i.e., ${\mathcal C} = ({\mathcal C}_\text{inner})^{V-1}$.
The inner loop can be pictured as a hop from one node to the next connected one.
The outer loop ensures that every node has been reached if it is at all reachable, because the shortest hop-distance between two nodes cannot be larger than $V-1$.

\subsubsection{Inner loop (${\mathcal C}_{\rm inner}$)}

The inner loop iterates over all edges $(i,j) \in \cal E$.
Because this operation acts over all edge failure configurations, where there exists at least one configuration where an edge has failed, we need to iterate in the most general way over all edges, and no particular iteration order is advantageous over another.
For specific graphs with obvious symmetries (e.g., linear graphs), certain iteration orders can indeed be more beneficial, but for a general graph this is not the case.

Without loss of generality, our starting point is edge $(i,j)$ that connects nodes $i$ and $j$. Consider now the following transformation:
\begin{align}
\ket{e_{i,j}}\ket{v_i, v_j} \rightarrow
\ket{e_{i,j}}\ket{v_i, v_j\oplus (v_{i}{e}_{i,j})\bar{v}_j}.
\end{align}
This transformation, where $\rightarrow$ here assumes suppression of any modulus-one phase in the output state,
induces a quantum-conditioned OR (qc-OR) operation in the Boolean states of the quantum state, i.e., for a given $v_i$, $v_j \mapsto v_i v_j + v_i + v_j$ if ${e}_{i,j} = 1$ and $v_j \mapsto v_j$ if ${e}_{i,j} = 0$.

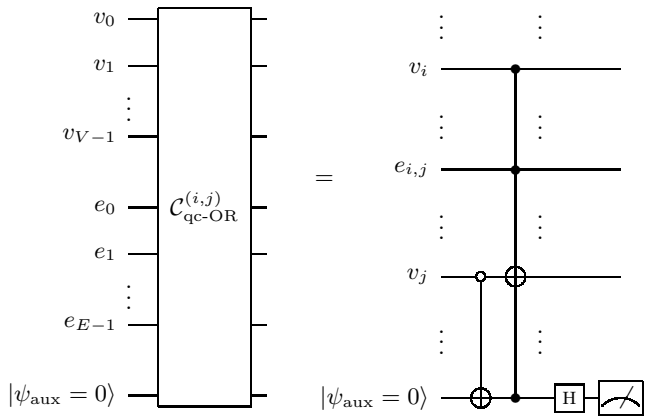
\begin{figure}\[
\hspace{8.5mm} 
\Qcircuit @C=0.6em @R=1.0em {
&&&& \lstick{v_0}  & \qw & \multigate{10}{\mathcal C_\text{qc-OR}^{(i,j)}} & \qw\\
&&&& \lstick{v_1}  & \qw & \ghost{\mathcal C_\text{qc-OR}^{(i,j)}} & \qw \\
&&&& \vdots        & & &  \\
&&&& \lstick{v_{V-1}} & \qw & \ghost{\mathcal C_\text{qc-OR}^{(i,j)}} & \qw \\
\\
&&&& \lstick{e_0}  & \qw & \ghost{\mathcal C_\text{qc-OR}^{(i,j)}} & \qw\\
&&&& \lstick{e_1}  & \qw & \ghost{\mathcal C_\text{qc-OR}^{(i,j)}} & \qw\\
&&&& \vdots        & & & \\
&&&& \lstick{e_{E-1}} & \qw & \ghost{\mathcal C_\text{qc-OR}^{(i,j)}} & \qw \\
\\
&&&& \lstick{\ket{\psi_\text{aux}=0}}  & \qw & \ghost{\mathcal C_\text{qc-OR}^{(i,j)}} & \qw \\
}
 \raisebox{-6.5em}{\hspace{6.5mm}=\hspace{6.5mm}}  
\Qcircuit @C=0.6em @R=1.93em {
&&&& \vdots & & & &  \vdots & & \\
&&&& \lstick{v_{i}}  &\qw & \qw & \ctrl{2}  & \qw & \qw & \qw \\
&&&& \vdots & & & &  \vdots & & \\
&&&& \lstick{e_{i,j}} &\qw & \qw & \ctrl{2} & \qw & \qw & \qw \\
&&&& \vdots & & & &  \vdots & & \\
&&&& \lstick{v_{j}} & \qw &  \ctrlo{2}& \targ & \qw & \qw & \qw  \\
&&&& \vdots & & & &  \vdots & & \\
&&&& \lstick{\ket{\psi_\text{aux}=0}}  &\qw & \targ & \ctrl{-2} & \qw & \gate{\hgate} & \meter  \\
}
\]
    \caption{\label{fig:outer-inner} Circuit diagram for the quantum-controlled OR step of the reachability operation. The circuit
    computes $v_i (1-v_j) e_{i,j} + v_j$ on the qubit labeled $v_j$ up to a modulus-one phase, where $v_i, v_j, e_{i,j}$ encode the reachability and failure states of the corresponding nodes and edge, respectively.}
\end{figure}

Note, by construction of our algorithm, there never is a superposition of the computational basis states of the node register, for a given edge configuration $e$.
In this way, the qc-OR operation, implemented according to the right-hand side of Fig.~\ref{fig:outer-inner} that leverages an in-circuit measurement,
induces the aforementioned transformation faithfully, and only introduces a potential sign flip in the coefficient of the output state.
This is critical for preserving the edge configuration probabilities $|f_e|^2$ [see Eq.~(\ref{eq:measurement})], especially as we apply multiple times the transformation in our quantum computer, leading to a moot, non-deterministic phase $(-1)^{\alpha_e}$ [see Eq.~(\ref{eq:outer-loop})].

Briefly, this sign flip may happen multiple times because each node $v_j$ may be visited several times within the inner-loop as multiple edges connect to $v_j$.
The total number of times the sign flipped is non-deterministic due to the in-circuit measurement of the ancilla qubit. It is further affected by the edge configuration $e$ and the particular order each edge is visited in the inner-loop.

\begin{figure}[!ht]
\[
\Qcircuit @C=0.6em @R=1.0em {
&&&&               &     &                            & \\
&&&& \lstick{v_0}  & \qw & \multigate{10}{\mathcal C} & \qw\\
&&&& \lstick{v_1}  & \qw & \ghost{\mathcal C} & \qw \\
&&&& \vdots        & & &  \\
&&&& \lstick{v_{V-1}} & \qw & \ghost{\mathcal C} & \qw \\
\\
&&&& \lstick{e_0}  & \qw & \ghost{\mathcal C} & \qw\\
&&&& \lstick{e_1}  & \qw & \ghost{\mathcal C} & \qw\\
&&&& \vdots        & & & \\
&&&& \lstick{e_{E-1}} & \qw & \ghost{\mathcal C} & \qw \\
\\
&&&& \lstick{\ket{\psi_\text{aux}=0}}  & \qw & \ghost{\mathcal C} & \qw \\
}
 \raisebox{-8.5em}{\hspace{7mm}=\hspace{7mm}}  
\Qcircuit @C=0.6em @R=1.0em {
&&&&               &     &  & \mbox{$\times \, V{-}1$} \\
&&&&   & \qw & \multigate{10}{\prod_{\varepsilon}{\mathcal C}_\text{inner}^{\varepsilon}} & \qw\\
&&&&   & \qw & \ghost{\prod_{\varepsilon}{\mathcal C}_\text{inner}^{\varepsilon}} & \qw \\
&&&& \vdots        & & &  \\
&&&&  & \qw & \ghost{\prod_{\varepsilon}{\mathcal C}_\text{inner}^{\varepsilon}} & \qw \\
\\
&&&&  & \qw & \ghost{\prod_{\varepsilon}{\mathcal C}_\text{inner}^{\varepsilon}} & \qw\\
&&&&  & \qw & \ghost{\prod_{\varepsilon}{\mathcal C}_\text{inner}^{\varepsilon}} & \qw\\
&&&& \vdots        & & & \\
&&&& & \qw & \ghost{\prod_{\varepsilon}{\mathcal C}_\text{inner}^{\varepsilon}} & \qw \\
\\
&&&&  & \qw & \ghost{\prod_{\varepsilon}{\mathcal C}_\text{inner}^{\varepsilon}} & \qw \\
}
\]
    \caption{\label{fig:EachNode} Circuit diagram for the reachability operator ${\mathcal C}$.
    ${\mathcal C}_\text{inner}$ is repeated $V-1$ times, which is defined as ${\mathcal C}_\text{inner}^{\varepsilon=(i,j)} = {\mathcal C}_\text{qc-OR}^{(i,j)}{\mathcal C}_\text{qc-OR}^{(j,i)}$ and is applied for every for each edge $\varepsilon \in {\mathcal E}$.}
\end{figure}
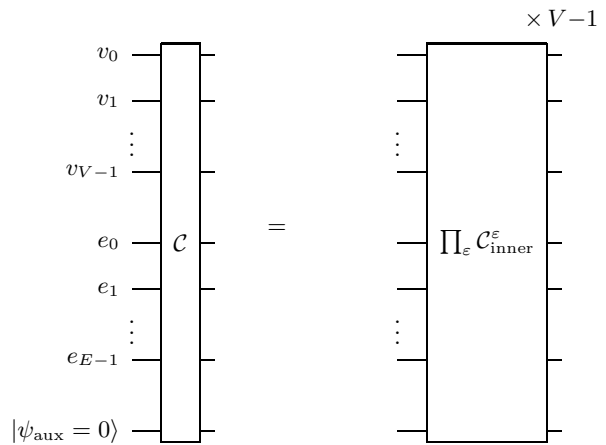

Equipped with the ability to induce a qc-OR operation, we introduce ${\mathcal C}_\text{inner}^{(i,j)} = {\mathcal C}^{(j,i)}_\text{qc-OR}{\mathcal C}^{(i,j)}_\text{qc-OR}$, which considers both directions of each edge in our undirected (or bidirectional) graph -- for directed graphs, we would consider the direction implied only.
We apply this symmetrized operation for every edge $(i,j)$ as our inner loop operations, as shown in Fig.~\ref{fig:EachNode}.
This concludes the inner-loop discussion, which visually speaking is equivalent to turning on every node if at least one neighboring node is on and the connection to it has not failed.

\subsubsection{Outer loop ($\mathcal{C}$)}
We repeat the inner-loop process described in detail above $V-1$ times, as shown in Fig.~\ref{fig:outer-inner}. 
This way, if it is possible for a given edge configuration $e$ to connect all nodes, the node register $n_e$ in \eq{outer-loop} reads all one after the reachability operator ${\mathcal C}={\mathcal C}_\text{inner}^{V-1}$ is applied.

Performing the inner-loop $V-1$ times has no impact on the discussed sign flips that are introduced by the in-circuit measurements in the inner loop.
After the entire reachability operator is performed (including the outer loop), the final sign is captured in the phase factor $(-1)^{\alpha_e}$ in (\ref{eq:outer-loop}). The modulus square $|f_e|^2$ in \eq{outer-loop} is still the probability of graph being in the edge-failure configuration $e$.

\subsection{Grover Oracle}
\label{sec:oracle}

To complete our algorithm, we need to estimate the probability that the node register reads an all-one output. A straightforward approach may be to directly measure the node register to see if it is in the all-one state, or to introduce a label qubit, initialized to $\ket{0}$, and apply a multiply-controlled inverter ($\notgate$) with the controls being the node qubits to see if the label qubit reads $\ket{1}$. Repeating this to collect statistics can then result in the estimation for the probability of the label qubit being in $\ket{1}$, which is the reliability metric $R$ defined in (\ref{eq:network-reliability}) that we are looking for.

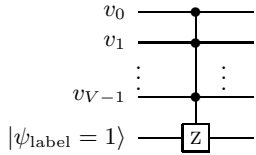
\begin{figure}
\[
\Qcircuit @C=0.6em @R=1.0em {
&&&& \lstick{v_{0}}   & \qw & \qw & \ctrl{4} & \qw & \qw & \qw \\
&&&& \lstick{v_{1}}   & \qw & \qw & \ctrl{3} & \qw & \qw & \qw \\
&&&& \vdots           & & & & \vdots & & \\
&&&& \lstick{v_{V-1}} & \qw & \qw & \ctrl{1} & \qw & \qw & \qw  \\
&&&& \lstick{\ket{\psi_\text{label}=1}}  &\qw & \qw & \gate{\zgate} & \qw & \qw & \qw  \\
}
\]
    \caption{\label{fig:GroverOracle} Circuit diagram for a Grover oracle. Considered is the case where we aim to estimate the exact reliability, i.e. ${\mathcal R} = 1$ [see Eq.~(\ref{eq:grover-oracle})]. The circuit implements $U_w$.}
\end{figure}

A more efficient approach is to consider a Grover oracle $U_w$ that acts on the node register, amplifying and helping identify various reliability or robustness metrics of interest for faster estimation. The oracle may be defined according to
\begin{equation}
    U_w \ket{v_0 v_1 \ldots v_{V-1}} = \begin{cases}   -\ket{v_0 v_1 \ldots v_{V-1}} \quad  &\sum_{l=0}^{V-1} v_l \geq T, \\
   +\ket{v_0 v_1 \ldots v_{V-1}} \quad  &\sum_{l=0}^{V-1} v_l < T,
    \end{cases} 
    \label{eq:grover-oracle}
\end{equation}
where $T$ defines which $T$-terminal reliability problem we are interested in.
For $T=V$, we have the all-terminal problem defined in (\ref{eq:network-reliability}).
Figure~\ref{fig:GroverOracle} shows how the quantum circuit for $U_w$ looks like for the all-terminal case.

\section{Complexity}
\label{sec:complexity}

\subsection{Gate Complexity}
 
In this subsection, we compute the gate complexity of our algorithm. The edge-register preparation takes $O(E)$ $\rygate$ gates. The node-register preparation takes $O(1)$ \notgate gate. The reachability operator takes $O(EV)$ qc-OR operations, each of which takes $O(1)$ \cnotgate and \tgate gates. Lastly, the Grover oracle has the gate complexity of $O(V)$ \cnotgate and \tgate gates. All combined forms a single Grover step.

For the Grover search to be successful, we need to repeat the Grover steps $O(1/\epsilon)$ times. This brings the total complexity to $O(EV/\epsilon)$.

\subsection{Space Complexity}
$O(E)$ qubits are needed to encode the edge failure and the node reachability. The only additional space we need is a single ancilla qubit, which can be reused every qc-OR operation. Furthermore, $O(V)$ ancilla qubits are required for the Grover oracle implementation. Therefore, the space complexity of our string-matching algorithm is $O(E)$. 

\section{Gate Counts}
\label{sec:gate-count}

In this section, we provide concrete gate counts. We count both \cnotgate gates and \tgate gates since, respectively, they are the widely-adopted, de-facto metrics for the quantum computational resource estimates for pre-fault tolerant and fault tolerant quantum computers. Note our gate-by-gate construction detailed above uses the standard gate set of \cnotgate and single-qubit Clifford$+$\tgate gates.

To compute the \cnotgate counts, we focus on the reachability operator and the Grover oracle. For the former, note the circuit shown in Fig.~\ref{fig:outer-inner} is applied at most $2E$ times, repeated $V-1$ times. Further, the triply-controlled \notgate gate can be optimized to a relative-phase Toffoli gate~\cite{maslov2016relative}, since our construction exhibits a ``don't-care'' behavior to the multiplicative phases introduced in the transformation, so long as the modulus is one (see \sec{reachability} for details). A relative-phase, triply-controlled \notgate consumes six \cnotgate gates. The circuit in Fig.~\ref{fig:outer-inner} therefore consumes seven \cnotgate gates in total, and a single call to the reachability operator consumes $14EV$ \cnotgate gates.

The Grover oracle for a $T$-terminal problem needs a single $T$-controlled \notgate, which can be constructed with $6T-12$ \cnotgate gates~\cite{maslov2016relative}. 

For the \tgate counts, an additional step of the initial edge register preparation needs to be considered. Within it, we need $E$ $\rygate$ gates, which we approximate using the well-known repeat-until-success technique~\cite{bocharov2015efficient}. Denoting the error budget for the initialization step as $\epsilon'$, and dividing the budget equally for individual $\rygate$ gates, we find the \tgate cost of the step to be $1.15\log_2(E/\epsilon')$. Note our amplitude estimation also incurs error $\epsilon$. To be consistent, we choose $\epsilon' \propto \epsilon$. For the reachability operator, each relative Toffoli gate with three controls consumes eight \tgate gates~\cite{maslov2016relative}. Applying them $2EV$ times results in $16EV$ \tgate gates. For the Grover oracle, each $T$-controlled \notgate gate consumes $8T-17$ \tgate gates~\cite{maslov2016relative}.

All together, the resulting, total \cnotgate and \tgate counts are
\begin{equation}
\begin{split}
&\#\cnotgate = (14EV + 6T - 12)\times 2/\epsilon, \\
&\#\tgate = (1.15\log_2(E/\epsilon)+16EV+8T-17) \times 2/\epsilon,
\end{split}
\label{eq:gate-estimate}
\end{equation}
where the factor of 2 stems from the amplitude amplification process~\cite{brassard2002quantum}.

\section{Conclusion}
\label{sec:conclusion}

We presented a quantum algorithm that computes the network reliability of an undirected graph---a well-known $\sharp$P-complete problem.
We discussed in detail its circuit-level implementation, and showed that our algorithm admits the gate complexity of $O(EV/\epsilon)$, which is a significant polynomial speed-up over the best classical algorithms currently known.
It is our hope that the quantum advantages like the one demonstrated herein will help enhance the reliability of future networks ranging from electric grids~\cite{LI201684} to telecommunication~\cite{103044,4781592} and transportation networks.

Here, we focused on the reliability calculation of undirected graphs.
There may be natural ways to extend this work to include directed graphs, which adds a level of complexity because reliability can no longer be rewritten as a root-node reachability problem with an arbitrarily chosen root node.
It would be interesting to see if further speed-up could be achieved by viewing the edges in terms of the adjacency matrix and reformulate the reachability question as a matrix problem.

In-circuit measurements are already supported~\cite{Honeywell2021} or planned~\cite{barnes2021assembly} in several quantum computer platforms.
Our approach exploits in-circuit measurements to create non-unitary operations---which we named qc-OR as it resembles a classical OR operation.
Our specific circuit implementation of the qc-OR operation enables us to keep the number of ancilla qubits to just one for the entire reachability operator, scale free in the problem size.
Our construction demonstrates the importance of non-unitary operations in harnessing the advantages of quantum computers for a broad class of computational problems.

We like to view this work as part of an wider effort to identify opportunities in new domains and problems where quantum computing could enable new possibilities.
In particular, we believe quantum computers have great potential to help with probabilistic problems that require explicit modelling of uncertainty.
Edge failure for the network reliability we studied here is just one obvious application.
Resource allocation~\cite{Wang2021-ResourcePGM} or routing~\cite{GhoshNgo07-RoutingProbGraph} problems in probabilistic graphs possibly are the areas where quantum computing could help push the limits beyond what is considered feasible today.

\section*{Author contributions}
S.P. and Y.N. designed the algorithm and prepared the manuscript. 

\section*{Competing interests:}
The authors declare that they have no competing interests. 

\section*{Data availability}
All data needed to evaluate the conclusions in the paper are present in the paper.
Additional data related to this paper may be requested from the authors. 
Correspondence and requests for material should be addressed to Y.N. (ynam@umd.edu) or S.P. (science@stefanpabst.name).

\bibliographystyle{naturemag}
\bibliography{citations}

\end{document}